
\magnification=\magstep 1
\baselineskip=16pt
\overfullrule=0pt

\def\ref#1{[#1]}
\rightline{HLRZ 43/92}
\bigskip\bigskip
\centerline{\bf ON THE THERMODYNAMICS OF GRANULAR MEDIA}
\bigskip\bigskip\bigskip\bigskip\smallskip
\centerline{{\bf H.J.~Herrmann}{\footnote{$^{\dagger }$}
{on leave from C.N.R.S. at SPhT, CEA Saclay, France}}}
\medskip
\centerline{H\"ochstleistungsrechenzentrum, KFA,}
\centerline{5170 J\"ulich, Germany}
\bigskip\bigskip\bigskip\vskip 3.5truecm
\noindent{\bf Abstract} \par
{\noindent
A thermodynamic formulation for moving granular material is proposed.
The fluctuations due to the constant flux and
dissipation of energy are controlled in a ``granular''
ensemble by a pressure $\wp$ (``compression'') which is conjugate to a
contact volume (``contactopy''). The corresponding response function
(``dissipativity'') describes how dissipation increases
with $\wp$ and should serve to identify the fluidization
transition and 1/f noise.
In the granular ensemble one can consider
the granular medium as a gas of elastically colliding particles
and define a ``granular'' temperature and other standard
thermodynamic quantities.
\vskip 2.0truecm\noindent
PACS: 05.70, 46.10 \hfill
\vfill\eject
Granular materials, like sand or powders,
subjected to an external force will locally perform
rather statistical motion due to the random
nature of the size and shape of grains and their contacts.
One example is the motion of sand on a vibrating plate, say a
loudspeaker\ref{1-6}. At sufficiently high frequency the individual
grains chaotically jump up and down forming a gas-like cloud
of colliding particles. Other examples are displacements
inside a shear-cell\ref{7-11} or flow down an
inclined chute\ref{11-15} where in addition to
a laminar flow with a well defined (average)
velocity profile one has Brownian-like motion of the
particles perpendicular to the flow direction.

The above observations have inspired several authors to
use thermodynamical concepts to describe
granular media. On one hand a ``granular temperature'' $T_{gr}$ has been
defined\ref{7,16,17} as $T_{gr} = \langle \vec v^2 \rangle
- \langle \vec v \rangle^2$, i.e. proportional to
the kinetic energy surplus with respect to the
global motion. This temperature has been determined numerically
as a function of various external parameters and material constants
and under certain conditions consistency
with experimental measurements was confirmed\ref{7}.
The drawback of the above definition is that it
is only thermodynamically justified if an
equipartition theorem exists which is not the case for
granular particles since they
dissipate energy at collisions.

Edwards and collaborators\ref{18-20}
have put forward an entirely different,
original idea: Based on the important observation
that granular materials do not conserve energy while the entropy
$S$ is well defined they proposed
to consider the volume $V$ to replace the internal energy
in the usual thermodynamic formalism. In this way a temperature-like
quantity $X = \partial V/\partial S$
which they called ``compactivity'' can be defined.
Although formally correct, this formalism is not easy to
justify from first principles. In particular, in many real
situations like on the vibrating table or on an
inclined plane, the volume is not well limited at large
heights. While Edwards's approach seems intuitively correct
for dense packings and the definition of $T_{gr}$ reasonable in the
limit of strong internal motions or weak dissipation
they fail in the corresponding opposite limit.

Purpose of the present note is to propose a different thermodynamic
approach to granular materials founded on similar principles
as equilibrium thermodynamics and which should at least
partially incorporate the intuitive pictures of previous work.

As opposed to usual thermodynamics of molecular gases the
elementary units of granular materials
are mesoscopic grains consisting of
many atoms each ($10^{15} - 10^{25}$). When these object
interact (collide) the Lennard-Jones potentials of the
individual atoms are unimportant and completely
different mechanisms must be considered. It is important that
on a microscopic scale the surface of the grains is rough.
Solid friction is the immediate consequence: When two touching grains
are at rest with respect to each other a finite force $F_s$ is needed
to trigger relative motion ({\it static friction}),
while moving against each other
a finite force $F_d$ is needed to maintain
the motion ({\it dynamic friction}). $F_d < F_s$ and both
only depend on the normal force and neither on the
velocity nor on the area of contact ({\it Coulomb law}).
No doubt, this picture is idealized and an entire discipline,
called tribology, has evolved to study solid friction in depth\ref{21}.
For our purpose it is, however, more convenient to remain on the
simple text-book level. The solid friction has the
crucial consequence that on the level of the elementary units,
namely the grains, the system does {\it not} conserve energy as
opposed to molecular thermodynamics. Another source of dissipation
can be plastic deformation of grains due to the normal force
acting at collisions.

If energy is not constantly pumped into a granular system it will
stop moving and fall into one of its static configurations.
Constant motion of the grains can only be produced when there is a
steady state of energy flux. We are, however, not interested
in this flux itself also because
it is difficult to measure experimentally.
We just want to describe the motion of the granular
particles in a similar way as one describes the motion of
molecules in a gas at a given temperature.
The presence of the energy flux and
the fact that on the level of the grains
on which we want to formulate a thermodynamics the
energy is locally dissipated (i.e. not conserved) will, however,
force us to introduce concepts beyond that of
usual equilibrium thermodynamics.

We will assume typical conditions for local ``equilibrium'':
Most experiments have velocity and density
gradients\ref{1-15} and in those cases only a subsystem
spatially small compared to the gradient should be considered.
An eventual energy flux into the system should distribute
the energy over it homogeneously. This constraint can also reduce
the size of the subsystem. Outside this subsystem a generalized
``heat bath'' is assumed. Spatial and temporal averages
should be exchangeable (``ergodicity''). We will in fact
in the following consider {\it temporal}
averaging for practical (numerical) purposes. The averaging
procedure can even be complicated\ref{4,6} by the existence
of density waves.

It is important to notice that the dissipated
energy is of course only lost on the mesoscopic level -
microscopically this energy will be transformed essentially into heat
and blown away by the surrounding air.
This gives us a reasonable starting point for the
formulation of an analogy to usual thermodynamics.
It seems natural to consider energy
conservation as the first ``thermodynamical principle'':
$$\Delta I = \Delta E_{int} + \Delta D \ \ \ .\eqno(1)$$
The internal energy $E_{int}$ is like in
traditional thermodynamics
the kinetic and potential energy of all the degrees
of freedom of the grains as elastic bodies
(translation, rotation, elasticity, etc).
$\Delta D$ is the energy dissipated in a given time and
$\Delta I$ is the energy that was pumped into the system
while $\Delta D$ was dissipated in order to maintain the
steady state.\footnote{$^{\dag}$}{
In contrast to traditional thermodynamics we have an energy
flux and the dissipated energy itself
increases with time. Therefore one can
formulate eq.~(1) alternatively as $J_I = \dot E_{int} + \dot D$
where $J_I$ is the energy flux into the
system. Although in some cases this description gives a more intuitive
physical picture we will prefer in the following to argue only
in terms of changes $\Delta D$ and $\Delta I$
during a fixed time interval.}
Usually $\Delta I$ is some kind of work
(gravity on the inclined plane, ${1\over 2} A \omega^2$ on the
loudspeaker, etc).
If one allows for changes in the volume of the system
then eq.~(1) will become
$\Delta I = \Delta E_{int} + \Delta D + \Delta W$
where $\Delta W$ is the work done to change the volume.

Let us give to the excess dissipated energy $\Delta {\cal D} =
\Delta D - \Delta I$ in the following the
nick-name ``dissipate''. We will deal with $\cal D$
in a similar way as one treats in usual thermodynamics the heat.
Like the heat, the dissipate is not a potential since it
depends on the process by which a given state is reached.
It does, however, not stem from the kinetic
energy of the particles as the heat in a molecular gas
but is due to collisions, i.e. two particles coming together,
touching and separating again.

The dissipated energy is proportional to the sum of
normal forces $f^i_n$ that push the particles
together during collision $i$. One can therefore express changes
in $\cal D$ as $$\delta {\cal D} = \wp \delta C\eqno(2)$$
where $\wp$ is an internal pressure acting at collisions
that we shall call ``compression''.\footnote{$^{\ddag}$}{
In real collisions also shear forces can contribute to dissipation
so that $\wp$ would then be a like a stress tensor. For that
reason we explicitely did not call it pressure.}
It can be defined as $\wp = \rho
\langle f^i_n / A_i \rangle$ where $f^i_n$ is the normal force
and $A_i$ the area of contact of collision $i$ and the average
is performed over all collisions.
$\rho$ is the density of collisions, defined as the number
of collisions per unit volume and unit time.
It is easy to determine $\wp$ numerically.
When the particles do not have collisions
the compression is zero and no energy is dissipated.
The quantity $C$, which we will in the following call
``contactopy'', in analogy to the entropy, has the dimension of
a volume (contact volume). It is defined as the conjugate
variable to the compression $\wp$.

The contactopy has contributions due to plasticity and due to dynamic
and static friction.
Let us in the following argue for a geometrical interpretation of $C$
and consider first the two contributions from friction.
The dynamic (or better kinetic) part of the contactopy is
proportional to $\sum_i A_i \ell_i$
where $\ell_i$ is the distance over which two solid
grains slip during collision $i$. Since
$\ell_i$ is given by the collision time multiplied by the
velocity of the particles
this part is proportional to the particle overlap volume $V_{ov}$
that one has (for technical reasons) in molecular dynamic
simulations\ref{4,6,10,15,22}. Apart from geometrical prefactors
the proportionality constant is the dimensionless dynamic
friction coefficient $\mu_d$,
i.e. a material constant of the grains.
The overlap volume $V_{ov}$ can be defined more precisely
as the sterically excluded volume that would arise if the centers
of mass of the particles follow the real trajectories but one
does not take into account the elasto-plastic deformation.
The static (or potential) contribution of friction
to the contactopy $C$ only comes into play when the elastic
(potential) energy of two unlocking particles that were sticking
is released. It depends on the penetration depth $d_i$ at
collision $i$ because this determines the amount of material that
will be compressed (or fragmented). Therefore
this second contribution is proportional to $\sum_i A_i d_i$,
i.e. also proportional to $V_{ov}$. The proportionality factor contains
the static friction coefficient $\mu_s$
and the Young modulus $Y$ of the grains.
The other contribution to the potential part of the contactopy comes
from plasticity and is proportional to the size of the plastic
zone, i.e. again to $V_{ov}$, when the material dependent
plastic yield force $F_p$ is reached.
The complex stick-slip mechanism\ref{10,23} that
is triggered between two rigid grains
by the interplay of static and dynamic friction makes it difficult to
determine precisely the resulting material dependent constant
$\gamma (\mu_d, \mu_s, F_p, Y)$ that following the simple
arguments given above relates via
$$C = {\gamma \over \rho '} V_{ov}\eqno(3)$$
the contactopy to the (average) overlap volume $V_{ov}$.
The division by the number $\rho '$ of collisions
per unit volume in eq.~(3) assures that $C$ is an extensive quantity.
The proportionality of eq.~(3) can be checked numerically
by measuring independently the input of energy and the average
internal pressure to get $C$ and summing up the
overlap volumina of the collisions to get $V_{ov}$.

The contactopy plays a central r\^ole here and in similarity to
the approach of Edwards\ref{18} can be interpreted as
a volume. But although it ressembles
the internal energy in having a kinetic and a potential contribution
it is analogous in our thermodynamic formalism to the entropy.
Because of eq.~(3) the contactopy represents a geometric
characterization of the system. This makes it likely to be
a total differential $dC$, i.e. independent on how the system
was driven into its state, while in contrast, the dissipate
depends on the work done on the system. Numerically this could
be checked by monitoring $V_{ov}$ but the ultimate test
should be experimental in analogy to Carnot's experiments.

The idea to define a potential for a dissipative system actually
dates back to Lord Rayleigh\ref {24} (``dissipation function'')
and has been worked out in detail by Prigogine and
collaborators\ref {25}. The contactopy is on one hand a concrete example
for such a potential on the other hand it does not only contain
the statistical aspects of an ``internal entropy production''\ref {25}
but has for physical reasons the dimension of a volume.
It would in fact be important to work out a statistical
interpretation of $C$ in the (space-time) phase space of the collision
events in analogy to Boltzmann's statistical definition of the entropy
in the space of all configurations of positions and velocities.

The ``equilibrium'' -
which is in fact a steady state driven by the energy flux -
can now be defined as the ensemble {\it minimizing} at fixed
$E_{int}$ the contactopy (instead of maximizing the entropy).
One can {\it postulate} an analog to the second law of thermodynamics
that any change of state at constant internal energy $E_{int}$ should
decrease the contactopy $$\Delta C \leq 0\ \ \ \ .\eqno(4)$$
Physically such a behaviour seems naturally be driven
by the elastic repulsion between colliding (overlapping) grains and
the tendency of the system to prefer many
smaller collisions to a few strong ones.
The third law of thermodynamics, namely that at zero
compression there is no overlap between grains, i.e.
vanishing contactopy, is less evident.
Numerical tests of the above statements
should be performed taking into account that as mentioned before
they are valid in (sub)systems into which the energy flow
allows for homogeneous dissipation.

As in usual thermodynamics one can now work in different ensembles.
One can fix either the compression $\wp$ which we shall call
the ``granular ensemble'' or the contactopy
(let's call it ``atomistic ensemble'').
Since in practice (experimentally and numerically), however,
the later case is difficult to
implement we will in the following usually
consider the granular ensemble.
On top of this we can build up the traditional body of
thermodynamics as if the grains were a gas of particles
interacting elastically. We can fix or free the number
$N$ of particles, define a ``granular'' temperature $T_g$ and entropy
$S$ or impose to the system either an external volume $V$ or an
external pressure $p$. A novelty for granular media is that
one could also impose an external shear $\tau$ or its
``conjugate'', the dilatancy $V_d$\ref{26}.

A granular potential $G_r$ can be defined as
$$G_r = E_{int} + \wp C\eqno(5)$$
which depends on $\wp$ and the extensive variables $N,V,S$ and $V_d$.
An immediate consequence of eq.~(4) is that at constant
compression $\wp$ the equilibrium is given by the minimum of $G_r$.
The analog to the specific heat might be called
``dissipativity'' $\kappa$ defined as
$$\kappa = {\partial {\cal D}\over \partial \wp} = \wp
{\partial C\over \partial \wp}\ \ \ .\eqno(6)$$
This is a new quantity characterizing the granular medium
which measures how much more energy can be dissipated
if the compression is increased.
It could be measured directly by numerically evaluating the
derivative of eq.~(6) or through the fluctuations of the energy
in a thermally closed system, i.e. surrounded by (infinitely)
heavy and stiff walls.
The dissipativity $\kappa$ should be positive and go to zero for
$\wp \rightarrow 0$ and $\wp \rightarrow \infty$.
Interesting for practical purposes is that
$\kappa$ contains through the $\gamma$ of
eq.~(3) the material dependent properties concerning friction,
among others also the stick-slip mechanism between grains.
If a fluctuation-dissipation theorem
for the response function $\kappa$ is valid
then one might identify $1/f$ noise\ref{23,27}
from its frequency dependence
over time scales proportional to the size of the grains or even
over larger time scales when collective phenomena like
arching or bridge-collapsing\ref{8,9} come into play.

We know that there exists a
``fluidization transition'' in granular media
between a regime of block motion at low energy flow to a
gas-like collisional regime at high energy flow\ref{3,4}.
This transition could be driven by changing $\wp$: For small
$\wp$ the potential part of $C$ dominates (block motion) and
for large $\wp$ the kinetic part of $C$ is relevant (collisional
motion). It seems likely that the transition point is given
by a singularity of the dissipativity $\kappa$. This could be
checked experimentally and numerically.

In the granular ensemble a ``dissipate'' bath (instead of a
heat bath) is coupled to the system and consequently the
internal energy $E_{int}$ of the granular material
is a fluctuating quantity. In order to give Boltzmann's statistical
interpretation to the entropy $S$ it is therefore conceptually better
to work in the atomistic ensemble: A ``state'' is given by the
positions, orientations, linear and angular velocities of the
grains as rigid bodies. In fact the entropy is well-defined as noted
already in ref.~18. A reasonable definition for a ``granular''
temperature $T_g$ would then be:
$$T_g = \biggl({\partial G_r\over \partial S}
\biggr)_{\wp}\ \ \ .\eqno(7)$$
Note that it is similar but not identical to the granular
temperature defined by previous authors\ref {7,16,17}.
$T_g$ is the variable that controls the granular canonical
ensemble with the granular free energy $F_g$ as potential,
defined as the Legendre transformation of the granular potential:
$F_g = G_r - T_g S$.
In equilibrium $F_g$ should have a minimum.
The usual Boltzmann distribution determines the statistical
weights of the states in this ensemble.
As in the case of the usual temperature one can measure $T_g$ by
monitoring the exchange of internal energy between a subsystem
and its heat bath which should obey this Boltzmann distribution.
Also a direct measurement of $T_g$ by changing an external pressure
(see eg.~(17.1) of ref.~28) should be possible.
A specific heat can be defined as a derivation with respect to $T_g$.

Experimentally $\wp$ and $T_g$ are independent control
parameters of the system: Since $T_g$ increases with the kinetic
energy of the particles it is essentially controlled by the
amount $\Delta I$ of energy that is fed into the system per unit time.
The compression $\wp$ or better the quantity
$\wp \over T_g$ also depends on
the density of collisions and can therefore increase by fragmenting
the grains into smaller pieces. (Note that when a given grain is
split into eight pieces, the cross section
of each individual piece decreases by a factor four, so that
$\wp$ will increase by two.) One can therefore, by changing the
grain diameters and $\Delta I$ modify $\wp \over T_g $ and $T_g$
independently and can therefore consider phase diagrams in the
$\wp \over T_g$ - $T_g$ plane.

As already mentioned one can also go to other ensembles by
changing variables via further Legendre transformations. One can
liberate the number of grains and introduce a granular
grandcanonical potential controlled by a chemical potential. One can
also fix an external pressure and calculate the average volume.
The fact that both volume and contactopy will then be conjugate
to the pressure could explain why in granular media
one finds in equilibrium macroscopic density fluctuations\ref{29}
as opposed to usual fluids or gases.
Of practical interest is also to fix an external shear and
measure the average expansion, i.e. the dilatancy\ref{8,9,26}.

It is useful to note that in the case when friction and plasticity
vanish the system does not dissipate energy
anymore, the contactopy will be zero and $G_r = E_{int}$.
In that case the atomistic and granular ensembles are identical
and classical thermodynamics is immediately recovered. Our formalism
is therefore a genuine generalization of equilibrium thermodynamics.

We have described within a thermodynamic formalism the fluctuations
arising from the constant flux and dissipation of energy that drives
a granular material's kinematic behaviour. By separating the
dissipative degree's of freedom (friction and plasticity) from
the conservative ones (translation, rotation, elasticity) we
define a ``granular ensemble'' coupled to a ``dissipate bath''
which is in fact the one in which experimental and numerical
measurements are usually performed. We introduce a potential that
we call ``contactopy'' and argue that it is proportional to the steric
overlap volume of the collisions which the
particles would have had per time unit if while
following the real trajectories they had no elasto-plastic deformation.
It would be interesting to give also a statistical interpretation
to the contactopy in order to define it as a
dissipative potential\ref {25}. The fluctuating
internal energy is replaced by a granular potential controlled
by an intensive variable that we call ``compression'', which is
conjugate to the contactopy.
Going into a granular canonical ensemble we define a
``granular temperature'' similar
to the one defined previously\ref{16,17}.
We propose various numerical and experimental tests for the assumptions
that we have made in our theory and suggest that a frequency dependent
``dissipativity'' should characterize the stick-slip
behaviour of the material and the transition to fluidization.
\bigskip
I thank R. Balian, J. Kert\'esz, M. Kiessling, J. Lebowitz,
C. Moukarzel, S. Roux, D. Stauffer, D. Wolf, R.K.P. Zia and in
particular H.G. Herrmann for encouraging discussions.
\bigskip\bigskip\bigskip
\centerline {\bf References}
\bigskip\medskip

\item{[1]}Faraday M., Phil. Trans. R. Soc. London {\bf 52}
(1831) 299; Walker J., Sci. Am. {\bf 247} (1982) 167;
 Dinkelacker F., H\" ubler A. and  L\" uscher E., Biol. Cybern.
     {\bf 56}, (1987) 51
\item{[2]}Evesque P. and Rajchenbach J., Phys.~Rev.~Lett. {\bf62},
    (1989) 44; C.~R. Acad. Sci. Ser.~2 {\bf 307} (1988) 1;
     {\bf307}, (1988) 223;
Laroche C., Douady S.  Fauve S., J. de Physique
 {\bf 50} (1989) 699;
Evesque P., J. Physique {\bf 51} (1990) 697;
Rajchenbach J, Europhys. Lett. {\bf 16} (1991) 149;
Cl\'ement E., Duran J. and Rajchenbach J., Phys. Rev. Lett.
{\bf 69} (1992) 1189
\item{[3]} Cl\'ement E. and Rajchenbach J.,
Europhys. Lett. {\bf16} (1991) 133
\item{[4]} Gallas J.A.C., Herrmann  H.J and Soko\l owski S.,
      Physica A, {\bf 189} (1992) 437
\item{[5]} Zik O. and Stavans J., Europhys.
Lett. {\bf 16} (1991) 255; Mehta A. and Barker G.C., Phys. Rev.
Lett. {\bf67} (1991) 394
\item{[6]} Gallas J.A.C., Herrmann  H.J and Soko\l owski S.,
      Phys. Rev. Lett. {\bf 69} (1992) 1371;
      Tagushi Y.-h., Phys. Rev. Lett. {\bf 69} (1992) 1367
\item{[7]}Campbell C.S. and Brennen C. E.,
J. Fluid Mech. {\bf 151} (1985) 167;
Campbell C.S., Annu. Rev. Fluid Mech. {\bf 22} (1990) 57;
Hanes D. M. and Inman D. L., J. Fluid Mech. {\bf 150} (1985) 357;
Walton O.R. and Braun R.L., J. Rheol. {\bf 30} (1986) 949
\item{[8]} Bashir Y.M. and Goddard J.D., J. Rheol. {\bf 35} (1991) 849
\item{[9]} Bagnold R. A., Proc. Roy. Soc. London A {\bf 295} (1966) 219
\item{[10]}Thompson P. A. and Grest G. S., Phys. Rev. Lett.
{\bf 67} (1991) 1751
\item{[11]} Jaeger H.M., Liu C.-H., Nagel S.R. and Witten T.A.,
Europhys. Lett. {\bf11} (1990) 619
\item{[12]} Drake T. G., J. Geophys. Res. {\bf 95} (1990) 8681
\item{[13]} Johnson P. C. and Jackson R., J. Fluid Mech.
{\bf 150} (1985) 67
\item{[14]} Savage S. B., J. Fluid Mech. {\bf 92} (1979) 53;
Savage S.B. and Hutter K., J. Fluid Mech. {\bf 199} (1989) 177;
Savage S.B., in {\it Disorder and Granular Media} ed. D. Bideau
(North-Holland, Amsterdam, 1992)
\item{[15]} P\"oschel T., preprint HLRZ 22/92
\item{[16]} Ogawa S., {\it Proc. of US-Japan Symp. on Continuum Mechanics
and Statistical Approaches to the Mechanics of Granular Media},
eds. Cowin S.C. and Satake M. (Gakujutsu Bunken Fukyu-kai,
1978), p.208
\item{[17]}
Jenkins, J.T. and Savage, S.B., J. Fluid Mech. {\bf 130} (1983) 186;
Jaeger H.M., Lui C.-h. and Nagel S.R., Phys. Rev. Lett.
{\bf 62} (1989) 40;
Jaeger H.M. and Nagel S.R., Science {\bf 255} (1992) 1523
\item{[18]} S.F. Edwards and R.B.S. Oakeshott,
Physica A {\bf 157}  (1989) 1080
\item{[19]} Edwards S. F., J. Stat. Phys. {\bf 62} (1991) 889
\item{[20]}
 Mehta A. and  Edwards S. F., Physica A {\bf 157} (1989) 1091
\item{[21]} Johnson K.L., {\it Contact Mechanics} (Cambridge Univ.
Press, 1989)
\item{[22]} Haff P.K. and  Werner B.T., Powder Techn.
     {\bf 48} (1986) 239; Cundall P.A. and Strack O.D.L.,
G\'eotechnique {\bf 29} (1979) 47
\item{[23]} Chen K., Bak P. and Obukhov S.P., Phys. Rev. A
{\bf 43} (1991) 625; Thompson P.A. and Robbins M.O., Science
{\bf 250} (1990) 792; Feder H.J.S. and Feder J., Phys. Rev. Lett.
{\bf 66} (1991) 2669;
Carlson J.M. and Langer J.S., Phys. Rev. Lett., {\bf 62} (1989) 2632
\item{[24]} Lord Rayleigh, Proc. Lond. Math Soc. {\bf 4} (1873) 357
\item{[25]} Prigogine I., Bull. Acad. Roy. Belg. Cl. Sci. {\bf 31}
(1954) 600; Glansdorff P. and Prigogine I., {\it Thermodynamic Theory
of Structure, Stability and Fluctuations}, (Wiley Interscience,
New York, 1971)
\item{[26]} Reynolds O., Phil. Mag. S. {\bf 20} (1885) 469
\item{[27]} Lui C.-h. and Jaeger S.R., Phys. Rev. Lett. {\bf 68}
(1992) 2301; Baxter G.W., PhD thesis
\item{[28]} Landau L.D. and E.M. Lifschitz, {\it Statistical Physics}
(Akademie Verlag, Berlin, 1987)
\item{[29]} Baxter G. W., Behringer R. P., Fagert T. and
Johnson G. A., Phys. Rev. Lett. {\bf 62} (1989) 2825;
T. P\"oschel, preprint
\bye